\documentclass[%
twocolumn,
superscriptaddress,
nofootinbib,
 amsmath,amssymb,
 aps,
pra,
]{revtex4-2}

\usepackage[utf8]{inputenc}

\usepackage{mathtools}
\usepackage{amsfonts}
\usepackage{mathrsfs}
\usepackage{bbm}
\usepackage{slashed}
\usepackage{upgreek}
\usepackage{graphicx}
\usepackage{dcolumn}
\usepackage{bm}
\usepackage{bbold}
\usepackage[colorlinks=true,linkcolor=blue,citecolor=blue,urlcolor=blue]{hyperref}% add hypertext capabilities
\usepackage[mathlines]{lineno}
\usepackage{xspace}
\usepackage{siunitx}
\usepackage[capitalise]{cleveref} 
\usepackage{xcolor}

\DeclareMathOperator{\tr}{tr} 
 
\DeclareMathOperator{\const}{const}
\DeclareMathOperator{\flow}{flow}
 
\renewcommand{\phi}{\upvarphi}
\renewcommand{\rho}{\uprho}

\DeclareMathOperator{\STr}{STr}

\DeclareMathOperator{\V}{{\mathrm V }}
\DeclareMathOperator{\W}{{\mathrm W}}
\DeclareMathOperator{\vv}{{\mathrm V }}
\DeclareMathOperator{\ww}{{\mathrm W}}

\begin{document}

\title{Superfluidity in multicomponent fermions via the functional renormalization group}% 

\author{Michal Hnati\v{c}}%
\email{hnatic@saske.sk}
\affiliation{ Faculty of Science, P. J. \v{S}af\'{a}rik University, Park Angelinum 9, 041 54 Ko\v{s}ice, Slovakia}%
\affiliation{Institute of Experimental Physics SAS, Watsonova 47, 040 01 Ko\v{s}ice, Slovakia}
\affiliation{Joint Institute for Nuclear Research, Joliot-Curie, 6, 141980 Dubna, Russia
}%

\author{Georgii Kalagov}
\email{kalagov@theor.jinr.ru}
\homepage{https://orcid.org/0000-0003-2208-0040}
\affiliation{Joint Institute for Nuclear Research, Joliot-Curie, 6, 141980 Dubna, Russia
}%

\begin{abstract}
We reveal the critical properties of the phase transition towards superfluid  order that has been proposed to occur in large spin fermionic systems. For this purpose, we consider the bosonic field theory for fluctuations of the complex skew-symmetric rank-2 tensor order parameter close to the transition. We then nonperturbatively
determine the scale dependence of the couplings of the theory by means of the functional renormalization group.  We established a fluctuation-induced first-order phase transition.  In the weak-coupling regime, the jump in the order parameter is small and a new phase occurs 
almost continuously, while in the strong one the discontinuity of the transition is well detectable. 

\end{abstract}

\maketitle

\section{\label{sec:level1}Introduction}

Fermionic systems capable of developing instability towards pair binding, under certain circumstances, reveal a striking symbiosis of interparticle interaction and the symmetry of internal degrees of freedom. Among such systems permitting a fruitful combination of experimental and theoretical research, ultracold atoms stand out.  The fine tunability of trapped atomic species allows for a unique simulating and  benchmarking  of many-body theory across different issues of physics. The internal geometry of the interaction in numerous systems  can be classified into  underlying symmetry groups. Particle states in the corresponding group representation are labeled with subscripts, such as spin, flavor, pseudospin, and thus shape a multicomponent system. A prosaic example is tied to $SU(2)$ symmetry, which covers the most widely studied spin-1/2 systems.  Enlarged $SU(n)$ symmetry arises, inter alia, in  fermionic atoms where the nuclear $I$ and electron spins are decoupled, leading to $n = 2 I+1$ internal states \cite{Wu2010,Cazalilla2014,Gorshkov2014}. Typical atomic species as $^{87}\mathrm{Sr}$ with $I=9/2$ and $^{173}\mathrm{Yb}$ with $I=5/2$ are available now in cold atom experiments \cite{Gaebler2010,Pagano2014,Scazza2014,Zhang2014,Taie2012,Krauser2012,Taie2010,Desalvo2010,Cappellini2014}. Large spin systems possess a rich internal structure and manifest thus exotic phases and phenomena such as novel magnetism \cite{Duine2005,Parsaei2009,Ho2015,Chen2005}, spin-mixtures \cite{Tey2010,Ravensbergen2020,Abeln2021}, and Fermi-liquid behavior \cite{Cheng2017,Yip2014,How2018}. Their superfluid properties and Cooper pairing have attracted unquenchable  interest \cite{Ho1999,He2006,Rapp2007,Wu2003,Zhang2021,Ozawa2010,Schlottmann2014,Honerkamp2004,Capponi2008,Capponi2007,Cherng2007}. The theoretical study also touches upon  $Sp(n)$-symmetric systems \cite{Ramires2016,Ramires2017,Kataoka2011}.

In the vicinity of the critical point, the microscopic symmetry of systems comes to the fore, and the long-range behavior is explored through the prism of  effective field models that respect the original symmetry. The renormalization concept allows a consistent consideration of critical fluctuations and groups their relevant contributions into anomalous dimensions of thermodynamic quantities. It has been implemented in a perturbative fashion for a tremendous number of physical models. The $SU(n)$-symmetric fermionic systems also did not stand aside.

The bosonised $SU(n)$ fermionic theory in the Cooper channel was explored using the $\varepsilon$-expansion in \cite{komarova2013,kalagov2016,Nalimov2020}. The order parameter of the superfluid phase transition is given by the skew-symmetric $n\times n$ complex matrix. The respective $\beta$-functions were evaluated up to six loops \cite{bednyakov2021}. No infrared attractive fixed points of the renormalization group flow were found for $n>2$. In addition, the running coupling constants go beyond the stability domain. Instability in the theory at this stage is usually an indication of a possible phase transition in the spirit of the Coleman-Weinberg mechanism.  The impact of a magnetic field, treated as a gauge one, was reported in \cite{Antonov2016} as well. It was shown that the type of the fixed points elucidated earlier remains unaltered. Despite the efficiency of the $\varepsilon$-expansion  and the Borel resummation technique based on Lipatov's asymptotics, even qualitative predictions obtained in this way should be taken carefully with regard to discontinuous phase transitions, as highlighted in \cite{Kleinert2006} in the case of superconductivity. Therefore, the movement towards nonperturbative analysis could be an additional support for the theoretical description  of such phenomena.  The functional renormalization group (FRG) \cite{dupuis2021,Debelhoir2016,berges2002} overcomes the obstacles of  perturbation theory and offers a systematic way to explore a wide range of critical phenomena associated  not only with continuous phase transition but also with the discontinuous one.

In this paper, we study phase transitions in the $SU(n)$ model of complex skew-symmetric tensor field by means of the FRG. Our aim  is to find  a discontinuous phase transition induced by the fluctuation corrections in the multicomponent case $n >2$ and track the development of a finite jump in the order parameter. We employ the well-known local potential approximation (LPA) to derive the desired FRG flow equation, which we solve numerically using two paradigms. Firstly, we seek a solution in the form of Taylor's expansion which gives rise to  coupling constants flow and also recovers the one-loop perturbative $\beta$-functions. Secondly, we consider the full flow equation as a partial differential one and store it on the grid to obtain a global solution showing the emergence of the first-order phase transition.

Section ~\ref{sec:model} outlines the bosonic field model of superfluid modes we consider, the phase transition pattern we focus on,  and theoretical framework and nonperturbative approximations we use to derive the renormalization group flow equation for the  effective scale dependent potential. Section ~\ref{sec:sol} presents numerical results we have obtained via both polynomial expansion of the flow and the global solution.  Section~\ref{sec:end} concludes and provides an outlook. In the Appendices, we compute the RG flow and point out the numerical methods that are used in the main text. 

\section{\label{sec:model}Model and methods}

\subsection{\label{sec:ef}Effective field theory}
The underlying thermal Lagrangian \cite{abrikosov1975,ho1998,kawaguchia2012,stamper2013}
\begin{equation}\label{eq:lagrangian}
 \mathcal{L} =\psi^{*}_{i}  \left(\partial_t - {\bm \nabla}^2 -\mu \right)\,\psi_{i} -\frac{\lambda}{2}
\left(\psi_i^*  \psi_i\right)^2
\end{equation} 
describes a system of identical fermions occupying different hyperfine states $i = 1,\dots, n$ at temperature $T$ with the chemical potential $\mu$;  the coupling strength $\lambda$ of the interparticle interaction is generally expressed in terms of the $s$-wave scattering length or adjustable trap parameters. The Grassmann functions $\psi_i$ are antiperiodic on the  ``imaginary-time'' interval $t \in [0, 1/T]$ (hereinafter $k_B = \hbar = 2 \,m = 1$, where $m$ -- the particle mass). The Lagrangian is defined with respect to an ultraviolet momentum cutoff $\Lambda$ limiting its applicability, often related, in electron systems, to the bandwidth or Debye frequency, or the van der Waals length in systems of neutral atoms.

To address the physics close to the superfluid/superconducting phase transition, the authors of \cite{komarova2013} introduced auxiliary bosonic modes $\Phi_{i j}$ of a paired state by the Hubbard–Stratonovich transformation. A finite expectation value of this field below the critical temperature $T_c$ corresponds to the anomalous composite operator $\langle \Phi_{i j} \rangle = \lambda \langle \psi_i \psi_j \rangle$, which can be identiﬁed with the order parameter of the transition considered. The rank-2 antisymmetric tensor $\Phi_{i j}$ transforms as $\Phi_{i j} \to \mathcal{U}_{i k} \mathcal{U}_{j l}  \Phi_{k l}$, where $\mathcal{U} \in \mathrm{U}(n)$.

In the present work, we use as a starting-point the model derived in \cite{komarova2013}. The respective Ginzburg-Landau effective action containing gradient and local terms reads 
\begin{align} \nonumber 
    S_\textrm{GL} = &\int d^d x \Bigl[ \tr({\bm \nabla} \Phi^{\dag}\, {\bm \nabla}\Phi) + m_0^2 \tr(\Phi^{\dag}\Phi) + g_{01} (\tr(\Phi^{\dag}\Phi))^2 \\[1ex] & +   g_{02} \tr(\Phi^{\dag}\Phi \Phi^{\dag}\Phi)\Bigr] ,
\label{eq:GL}\end{align}
the input coefficients $m_0^2$ and $g_{01}, g_{02}$ depend on the parameters of the microscopic fermionic model $\mu$, $\lambda$, $T$, and $\Lambda$ \cite{kalagov2016,Boettcher2018}. Although the bare value $g_{01}=0$, because no corresponding vertex is generated from fermion loops, it has been added to ensure multiplicative renormalizability of the theory. Such a term is generated in the course of renormalization in any case.   The classical potential term has to be bounded from below, which restricts the coupling space as $g_{02}  > 0$ and $n g_{01} +  g_{02} > 0$.  If the system develops a fermion pair condensate at the transition temperature at which the ``mass''  $m_0^2$ changes sing, the potential acquires a non-vanishing ground state $\langle \Phi \rangle = \phi_1 t_1$ of the form 
\begin{equation}
\label{vacuum1}
t_1 = \begin{pmatrix}
 0 & \mathbb{1}_{n/2} \\ 
-\mathbb{1}_{n/2} & 0
 \end{pmatrix},
\end{equation}
with a real magnitude $\phi_1$; thus, the symmetry group $\mathrm{U}(n)$ breaks down to the  symplectic one $\mathrm{USp}(n)$ consisting of unitary matrices $M$ that satisfy the condition $M^{T} t_1 M = t_1$.

\subsection{\label{sec:rg}Renormalization group treatment}
The framework employed in this study is the functional renormalization group (FRG) \cite{dupuis2021, berges2002, pawlowski2007}. The essence of the approach resides in the scale-dependent effective average action $\Gamma_k$, which comprises all the fluctuations of the field model beyond the momentum scale $k$. That is, $\Gamma_k$ is infrared regularized by the cut-off function $R_k$ added to the original action $S$ as a momentum-dependent mass term. The functional $\Gamma_k$ smoothly interpolates between the original microscopic action on the ultraviolet scale $\Gamma_{\Lambda} = S$ 
and the full free energy -- the generating functional of 1PI Green functions -- at the infrared limit $\Gamma_{k\to 0} = \Gamma$. Successive infinitesimal scale steps are tantamount to a continuous flow in functional space governed by the Wetterich equation
\begin{equation}
\label{eq:weq}
     \partial_k \Gamma_k = \frac{1}{2} \STr\left\{ (\Gamma_k^{(2)}+R_k)^{-1} \partial_k R_k  \right\},
\end{equation}
where the trace $\STr$ is evaluated over all degrees of freedom, and $\Gamma_k^{(2)}$ is given by the second functional derivative of $\Gamma_k$  with respect to the fields. For the computation of the flow equation, there is some freedom to choose the regulator $R_k(p) = p^2 \, r(p/k)$, where the dimensionless shape functions  $r(x)$  we use are given by the optimized cut-off
\begin{equation}
r(x) = \left(\frac{1}{x} - 1\right) \,\Theta(1-x).
\end{equation}
It provides an optimal choice within the approximation scheme, which will be employed below \cite{litim2000, *litim2001}.

\subsection{\label{sec:lpa}Local potential approximation for the matrix  model}
The following analysis draws on the truncated effective average action $\Gamma_k$, where only the effective potential carries a scale dependence, that is, the leading-order approximation of the derivative expansion
\begin{equation}
\label{eq:lpa}
    \Gamma_k = \int d^d x \left[  \tr({\bm \nabla}\Phi^{\dag} {\bm \nabla}\Phi) + U_k(\Phi,\Phi^{\dag}) \right].
\end{equation}
The advantage gained by truncation is that the functional flow, \cref{eq:weq}, is brought into equations amenable to subsequent analytical/numerical evaluation. The effective potential $U_k(\Phi, \Phi^{\dag})$ is an arbitrary function which preserves the $\mathrm{U}(n)$ symmetry of the theory and thus depends on fields through the respective group invariants. Insertion of the Ansatz, \cref{eq:lpa}, into \cref{eq:weq} finally gives the flow equation for the potential $U_k(\Phi, \Phi^{\dag})$. The derived flow is then integrated down in scale starting from some short distance scale $\Lambda$, and one obtains the thermodynamic potential $U_{k \to 0}(\Phi, \Phi^{\dag})$ that encompasses all the macroscopic properties of the system. The initial condition -- the ultra-violet value $U_{\Lambda}(\Phi, \Phi^{\dag})$ -- can be read off directly from the GL action, \cref{eq:GL}; this yields the quartic potential parameterized by two coupling constants and the mass term. In the spontaneously broken state, the potential can be cast in the form 
\begin{equation}
    \label{uvpotential}
    U_{\Lambda}(\Phi, \Phi^{\dag}) = - m_{\Lambda}^2 \, \rho_1 + \frac{g_{1 \Lambda}}{2} \, \rho_1^2 + g_{2\Lambda} \, \rho_2, \quad m_{\Lambda}^2 > 0,
\end{equation}
where the constants $m_{\Lambda}^2, \,g_{1 \Lambda}, \,g_{2 \Lambda}$ are  identified with the appropriate combinations of the bare parameters $g_{01}, g_{02}$. The group invariants we use here are defined as 
\begin{equation}
\rho_1 = \tr(\Phi^{\dag}\Phi), \quad \rho_a = \tr\Bigl(\Phi^{\dag}\Phi- \frac{\rho_1}{n}\Bigr)^a, \quad a \geq 2.
\end{equation}
For any matrix field of a given size $n$ there is a finite set of 
independent invariants, this can be straightforwardly inferred from the Cayley–Hamilton theorem. We are interested in the case of  symmetry breaking resulting in the appearance of the vacuum expectation value, \cref{vacuum1}, in which there is the only non-zero invariant $\rho_1$, while the other ones vanish; thus, around this configuration the running potential $U_k(\Phi,\Phi^{\dag})$ can be represented in the form of an expansion in the small  high order invariants $\rho_n$ with coefficients being functions of $\rho_1$. In this study, we use the following approximate expression:
\begin{equation}\label{eq:fulpotential}
U_k(\Phi, \Phi^{\dag}) \approx \V_k(\rho_1) + \W_k(\rho_1) \, \rho_2.
\end{equation}
Information on the phase transition behavior we consider is certainly contained in the first term $\V_k(\rho_1)$; however, the respective flow is not closed and incorporates the next coefficient $\W_k(\rho_1)$, the flow equation of which must be deduced as well. To do so, we extend the vacuum expectation value, \cref{vacuum1}, in the orthogonal direction  $t_2$
\begin{equation}
t_2   = \begin{pmatrix}
0 & 1 & 0 &  \dots & 0 \\ 
-1 & 0 & 0 & \dots & 0 \\ 
0& 0& 0&  \dots & 0 \\
\dots & \dots & \dots & \dots & 0 \\ 
0 & 0 & 0& \dots & 0
\end{pmatrix},
\end{equation}
with an infinitesimal magnitude $\phi_2 \to 0$, that is
$\langle \Phi \rangle = \phi_1 t_1 + \phi_2 t_2$, and obtain both nonvanishing invariants $\rho_1 = n \, \phi_1^2 + 2 \,\phi_2^2$ and $\rho_2 = 4\,\phi_1^2 \,\phi_2^2 + o(\phi_2^2)$. The next step is to calculate the Hessian $\Gamma^{(2)}_k$ (see Appendix~\ref{sec:mass} for a detailed calculation of the mass spectrum) in the extended background and expand both sides of the total flow equation, \cref{eq:weq}, in $\rho_2$. Eventually, if we follow this line, we arrive at the RG flow
\begin{widetext}
\begin{subequations}
\label{eq:rgeq} 
\begin{align} % UnSCALED
 \partial_k  \V_k(\rho_1) = \, & C_d\,  k^{d+1} \left\{  \frac{1}{k^2+\V_k' + 2 \rho_1 \V_k''}  + \frac{(n+1)(n-2)/2}{k^2+\V_k' + 4 \rho_1 \W_k/n} + \frac{n (n-1)/2}{k^2+\V_k'}  \right\},\\[1ex]  \nonumber
  \partial_k  \W_k(\rho_1) = \, & C_d\,  k^{d+1} \left\{ \frac{(n-2) \W_k^2}{(k^2+\V_k')^3} + 
 \frac{9 (n+2)(n-4) \W_k^2}{n (k^2+\V_k' + 4 \rho_1 \W_k/n)^3}  - \frac{n (n \W_k + 2 (n-1) \rho_1 \W_k')}{
 4 \rho_1 (k^2+\V_k')^2}    \right. \label{eq:rgeq2}  \\[1ex] 
  &-\frac{1}{(k^2+\V_k' + 2 \rho_1 \V_k'')^2} \left(2\,{ \W''_k}\,\rho_{{1}}+5\,{ \W_k'}+{\frac {\W_k}{\rho_{{1}}}} -\frac {n \V_k''}{2 \rho_{1}}+\,\frac { \left( n{\V_k''}+4\,{\W_k'}\,\rho_{{1}}+4\,\W_k \right) ^{2}}{ 2 \rho_{1} \left( n
{\V_k''}-2\,\W_k \right)}  \right) \\[1ex]  \nonumber
&+ \left. \frac{1}{(k^2+\V_k' + 4 \rho_1 \W_k/n)^2}  \left(
\frac{n^2+4}{4 \rho_1}\,{ \W_k}-\frac{n^2+6}{2}\,{ \W_k'} -\frac {n \V_k''}{2 \rho_{1}}+\,\frac { \left( n{\V_k''}+4\,{\W_k'}\,\rho_{{1}}+4\,\W_k \right) ^{2}}{ 2 \rho_{1} \left( n
{\V_k''}-2\,\W_k \right)} \right)\right\}.
\end{align}
\end{subequations}
\end{widetext}
    The present system is a central algebraic result obtained for the optimized IR regulator. The seemingly singular terms $\sim 1/\rho_1$ in \cref{eq:rgeq2} cancel each other out; so the total flow is continuous at the origin $\rho_1=0$.  Also, note that in the case $n=2$ the system decouples into two independent equations. The first one describes the evolution of the $O(2)$ symmetric potential, which approaches the fixed point, and the physical system tends to build up all sorts of singular scales with the available set of critical exponents \cite{kompaniets2017, batkovich2016, dePolsi2020,balog2019}. The analytical form of the flow enables one to compare the FRG findings with the presently known perturbative results.

\section{\label{sec:sol}Numerical analysis}

\subsection{\label{sec:gelmann}Flow in a set of polynomials}

\begin{figure*}[t!]
\includegraphics[width=0.32\textwidth]{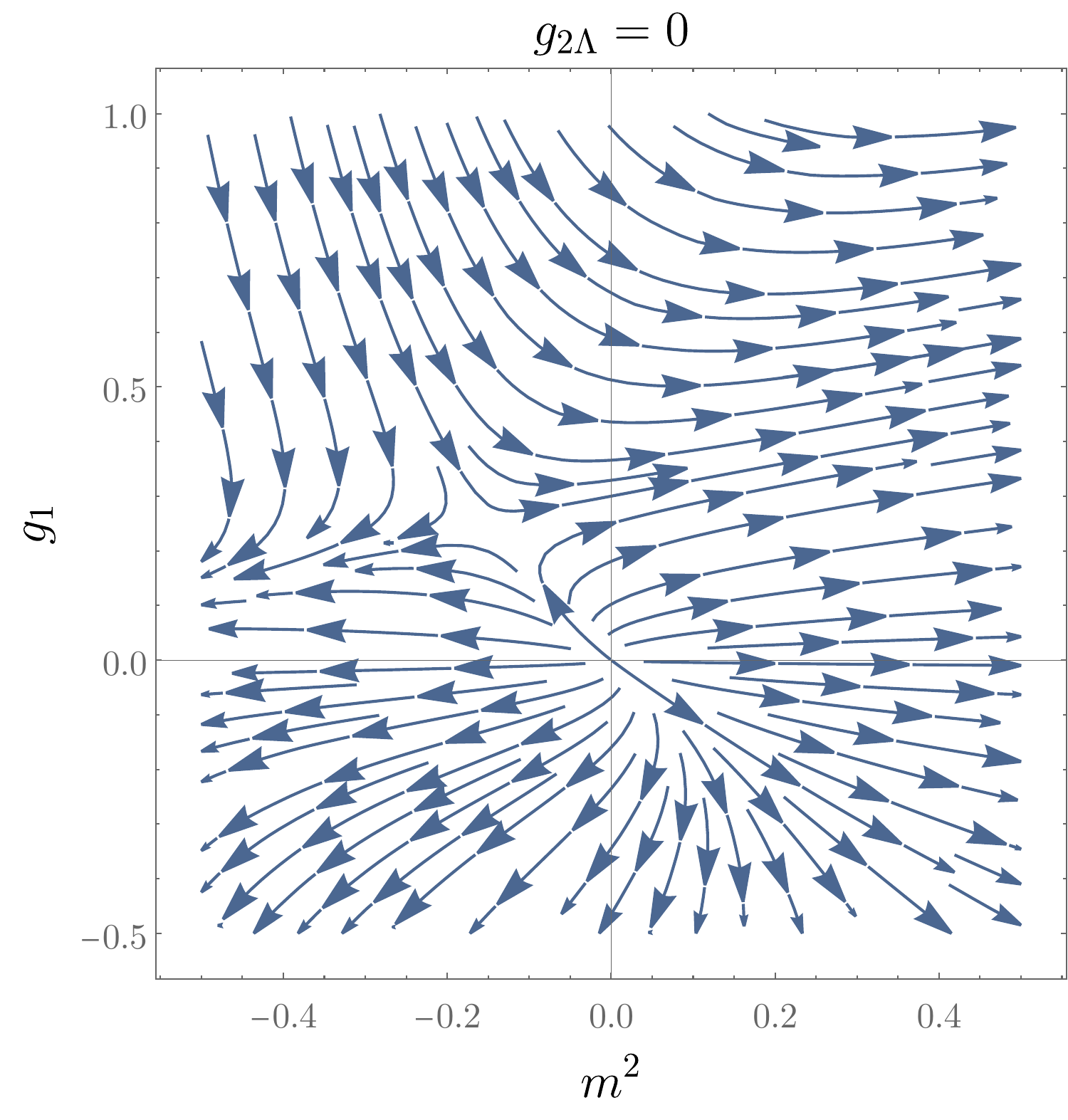}
\includegraphics[width=0.32\textwidth]{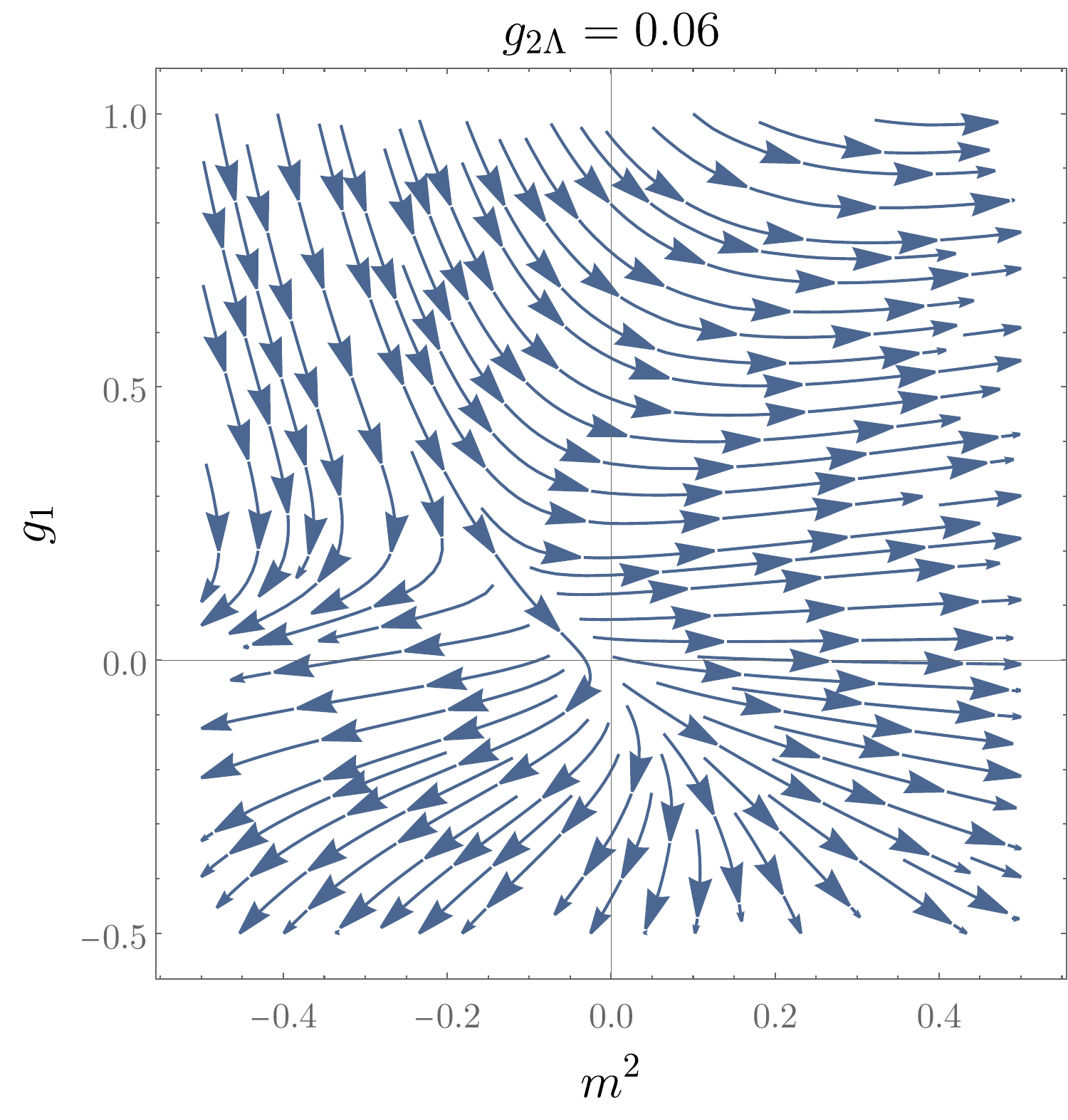}
\includegraphics[width=0.32\textwidth]{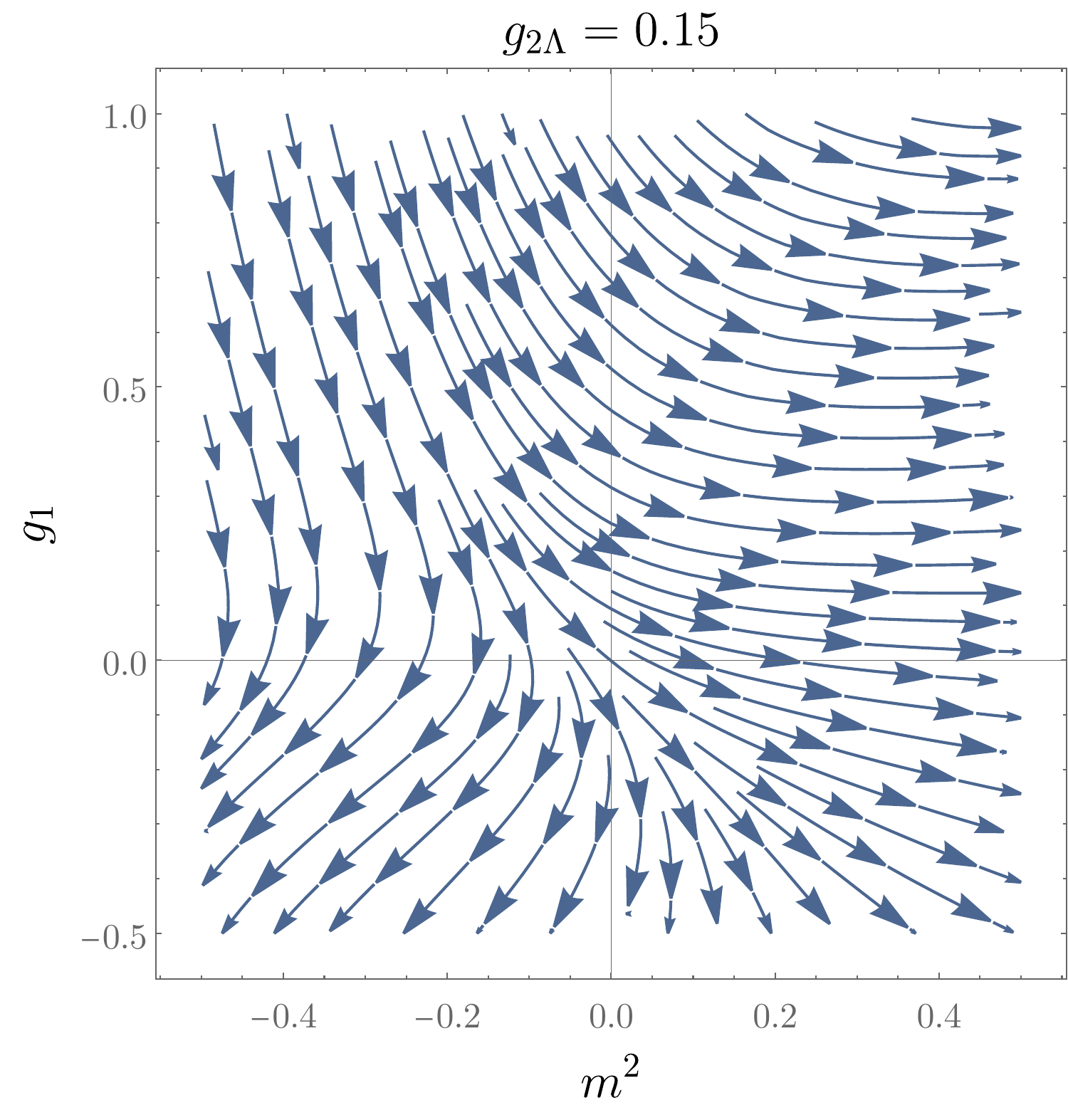}
\caption{\label{fig:phase} The RG flow in $(m^2, \,g_1)$-plane for a set of $g_{2\Lambda}$ values and $n=4$. The magnitude $g_{2\Lambda}$ is presented in units of $\Lambda$. }
\end{figure*}

To check the consistency of the derived equation with the $\varepsilon$-expansion, we need to expand functions $\vv_k, \ww_k$ as segments of a Taylor series in $\rho_1$ around the point $\rho_0$. There are several different ways how to realize an appropriate truncation scheme depending on the choices of the $\rho_0$ point: $\rho_0=0$ -- expansion above a phase transition; $\rho_0=\rho_0(k)$ -- expansion over running minimum of the potential $\vv_k$ and  $\rho_0=\const$ -- expansion over a point shifted from the origin  that is the minimum of the infrared potential $\vv_{k=0}$, which is however unknown a priori; therefore the magnitude $\rho_0$ can be tuned to attain a better convergence of the expansion \cite{Aoki1998}. Here we confine ourselves with the first case enough to derive one-loop results within the $\varepsilon$-expansion; thus, we set
\begin{equation}
\label{eq:potentials}
\begin{split}
     &\V_k(\rho_1) =\V_k(0) + m_k^2 \, \rho_1 + \frac{g_{1 k}}{2!}\, \rho_1^2, \\[1ex] % + \frac{g_{3 k}}{3!}\, \rho_1^3+\dots, \\[1ex]
     &\W_k(\rho_1) = g_{2 k}. %  + g_{4 k}\, \rho_1 + \dots.
\end{split}
\end{equation}
Substituting this expansion into the system, \cref{eq:rgeq}, and equating the successive coefficients of the powers of $\rho_1$, we obtain the system

\begin{widetext}
\begin{subequations}
\label{eq:gellmann}
\begin{align}
    \partial_k g_{1 k} &=   A_3 \left\{ 2 \, \left(1-\frac{1}{n}+\frac{8}{n^2}\right) \, g_{1 k}^2 +8\,  \left(1-\frac{1}{n}-\frac{2}{n^2}\right)  \left(  \,g_{1 k} \,g_{2 k}+2 \, g_{2 k}^2\right) \right\},\label{eq:g1}\\
    \partial_k g_{2 k} & =   A_3 \left\{  4\, \left(2 -\frac{5}{n}-\frac{12}{n^2}\right) \, g_{2 k}^2 + \frac{24}{n^2} \,g_{1 k} \,g_{2 k}  \right\},\label{eq:g2}\\
    \partial_k m_k^2 &=  - A_2 \left\{ \left(1-\frac{1}{n}+\frac{2}{n^2}\right) \, g_{1 k} +2\,\left(1-\frac{1}{n}-\frac{2}{n^2}\right) \, g_{2 k} \right\},\label{eq:m}
\end{align}
\end{subequations}
\end{widetext}
where $A_s \equiv  C_d\,  k^{d+1}/(k^2+m_k^2)^s $. Introducing the dimensionless charges $g_{1, 2} = g_{1, 2}(t)$ and $m = m(t)$, according to $g_{1, 2 \, k} = g_{1, 2} \,k^{4-d}$ and
$m_k^2 = k^2 m^2 $, where the RG time $t =  \ln(k/\Lambda)$, we obtain exactly the same $\beta$-functions as in \cite{kalagov2016, bednyakov2021} formed by keeping only the leading term in $\varepsilon = d - 4$ and defined as $\beta_{1, 2} = \partial g_{1, 2}/\partial  t$ \footnote{To obtain the same expressions, one needs to perform additional change of variables $g_1 \to 4 \pi^2 \, (g_1 + g_2/n)$, $g_2 \to  2 \pi^2 g_2$, and omit the $m^2$ term in the denominators of $A_s$. }. 

Figure \ref{fig:phase} demonstrates distinct 2D phase portraits of the RG flow, \cref{eq:gellmann}, for different choices of the $g_{2 \Lambda}$ magnitude. For the simplest case $g_{2 \Lambda} = 0$,  the
equation (\ref{eq:g2}) exhibits only the trivial solution $g_{2 k} = 0$ and consequently the system possesses  global $O(n^2 - n)$ symmetry.  This can be straightforwardly inferred by expansion of the $\Phi$ field  in terms of the $SO(n)$ generators with the $n^2 - n$ real coefficients, taking into account the absence of the $\tr(\Phi^{\dag}\Phi \Phi^{\dag}\Phi)$ vertex on all scales. The respective figure depicts the familiar behavior of the RG flow in the vicinity of two fixed points. One of them is infrared attractive and gives rise to anomalous scaling of correlation functions and thermodynamic quantities. The effect from a finite value of $g_{2 \Lambda}$  is to considerably destroy this flow pattern. No physically meaningful stable fixed point has been found. The same picture was obtained by the authors \cite{komarova2013} who, to our knowledge, were the first to point out this feature exploited in the context of multicomponent fermions through the one-loop renormalization group calculation within the $\varepsilon$-expansion. Subsequently, the perturbative analysis has been extended up to  the five-loop approximation in \cite{kalagov2016}. Recently, in \cite{Nalimov2020}, the convergent (unlike the $\varepsilon$-expansion) perturbative scheme has been proposed to overcome paradigmatic difficulty associated with the factorial growth of expansion coefficients. A conclusion to be drawn from these works is that the  flow pattern is robust against higher order contributions. Although in \cite{bednyakov2021} the six-loop $\beta$-functions were published for the general matrix $\phi^4$-type models, there seems no doubt that new corrections cannot in any way alter the already elucidated perturbative picture. A common, if somewhat inaccurate,  interpretation of this state of affairs, specialized to the cases of scaling phenomenon, is that the system does not undergo a continuous phase transition, while it may (and usually does) manifest the first-order phase transition behavior. The  pattern presented in \cref{fig:phase} shows that for strong enough $g_{2 \Lambda}$  even a system with initially positive $g_1$, i.e., it is in a physical region, may eventually ﬂow towards the ``unphysical'' one where $g_1 < 0$; thus, the potential is unstable and, in the absence of the sextic term $\rho_1^3$, does not describe a physical system. The corresponding flow for higher order monomials in \cref{eq:potentials} can be easily obtained from the general system, \cref{eq:rgeq}, and it is not presented in this manuscript, we will instead carry out a global analysis of the system,  \cref{eq:rgeq}, as a whole in the section below.  

Note that in \cite{kalagov2016}, sextic vertices like $(\Phi^{\dag}\Phi)^3$ 
were directly incorporated into the bare action, \cref{eq:GL}. From the renormalization theory perspective, these terms generate an infinite set of ultraviolet divergent diagrams which cannot be treated within the consistent iterative subtraction scheme. Therefore, they were considered to be composite operators in the course of analysis, namely, Green functions with the only insertion of the sextic vertices were renormalized. The justification for doing so rests explicitly on the assumption that the respective coupling constants remain small in the infrared region.  The  renormalization group functions for sextic composite operators can be extracted from \cref{eq:rgeq} except for one of them $\tr(\Phi^{\dag}\Phi)^3$ which requires the $\rho_3$ invariant to be involved in the approximation, \cref{eq:fulpotential}. It turns out, not surprisingly with hindsight, that a similar situation arose in the work devoted to the spontaneous breaking of chiral symmetry of QCD \cite{Fejos2015}. It has been shown numerically by solution of the flow in the $\rho_3$ direction that it does not even get any observable change of the final results. Thus, there is a sensible reason to believe that the truncation, \cref{eq:fulpotential}, is quite appropriate here. To conclude this section, it is necessary to stress that the above picture is qualitatively valid for all even values $n > 2$.

\subsection{\label{sec:global}Global flow}
\begin{figure*}[t!]
\includegraphics[width=0.48\textwidth]{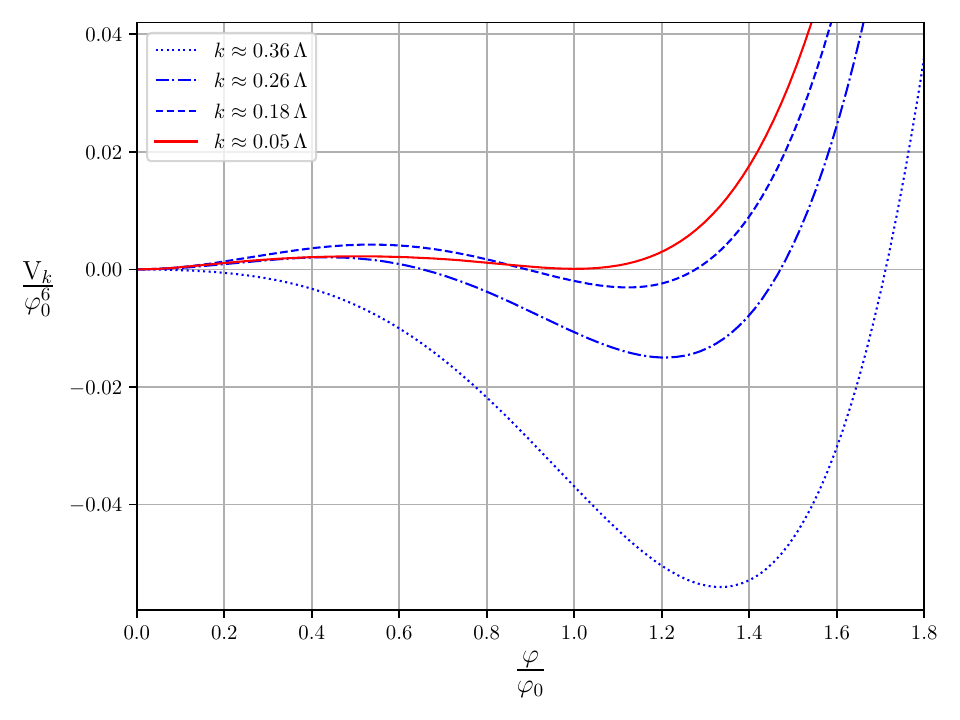}
\includegraphics[width=0.48\textwidth]{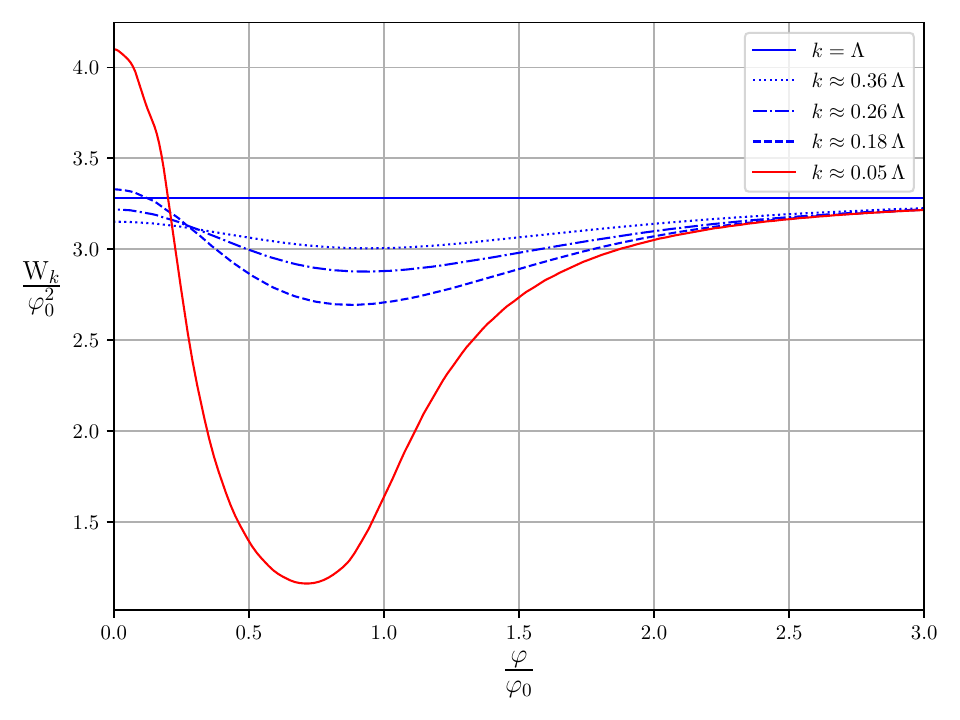}
\caption{\label{fig:evolution}The RG flow, \cref{eq:rgeq}, moves out from the origin $\vv_{\Lambda}, \ww_{\Lambda}$ towards the infrared as the scale $k$ decreases at the critical value $m^2_\Lambda = m^2_c$. The graphs are depicted in units of $\varphi_0$ for $n=4$. The straight red line refers to the infrared potentials obtained from the flow stopped at the scale $k  = k_{IR} \approx 0.05 \Lambda$. The quantity $\ww_k$ starting from a constant at $k = \Lambda$ develops a nontrivial shape. }   

\end{figure*}
To obtain the global solution of the system \cref{eq:rgeq}, the functions $\vv_k$ and $\ww_k$ were stored on the Chebyshev-Lobatto grid within the pseudo-spectral approach (which is set out in Appendix~\ref{sec:numer}). To propagate the solution between the scales $k$ and $k + \Delta k$, one employs the adaptive Runge-Kutta-Fehlberg scheme ensuring stabilization of the numerical procedure. Note, there is a rich variety of other numerical schemes, among them a promising role is played by a toolbox of numerical fluid dynamics. For instance, the discontinuous Galerkin method allows access to non-analyticities  (akin, for example, to shock waves in fluids) in the solution, which are intrinsic to issues of phase transitions. A comprehensive and detail discussion of this subject as applied  to zero-dimensional field models can be found in \cite{Wink2021}.

The endemic characteristic of the Legendre transformation is convexity of the full effective action $\Gamma$, or the free energy $\vv(\varphi)$ in a uniform system we are handling. In the thermodynamic limit, it may become not strictly convex and hence has a plateau in its shape, where the nontrivial solution of the state equation $\partial \vv(\varphi) /\partial \varphi = 0$ is degenerate because any value of $\varphi$ in this flat segment  $|\varphi| \leq \varphi_0$  obeys the state equation, while  the endpoints $\pm \varphi_0$  determine the finite superfluid density. This picture is not exactly fulfilled under the circumstances most often met in real computations based on diverse truncations; however, improving the accuracy of the approximation scheme actually reinforces the tendency to flattening the inner region  $|\varphi| \leq \varphi_0$. The coarse-grained free energy $\vv_k(\varphi)$ is not the Legendre transformation and does not therefore inherit convexity for a nonzero value of $k$.  
It also can roughly be considered as being associated with  a system confined in a box of finite size $\sim 1/k$; thus, no singularity emerges in the free energy.  

\begin{figure}[t!]
\includegraphics[width=0.48\textwidth]{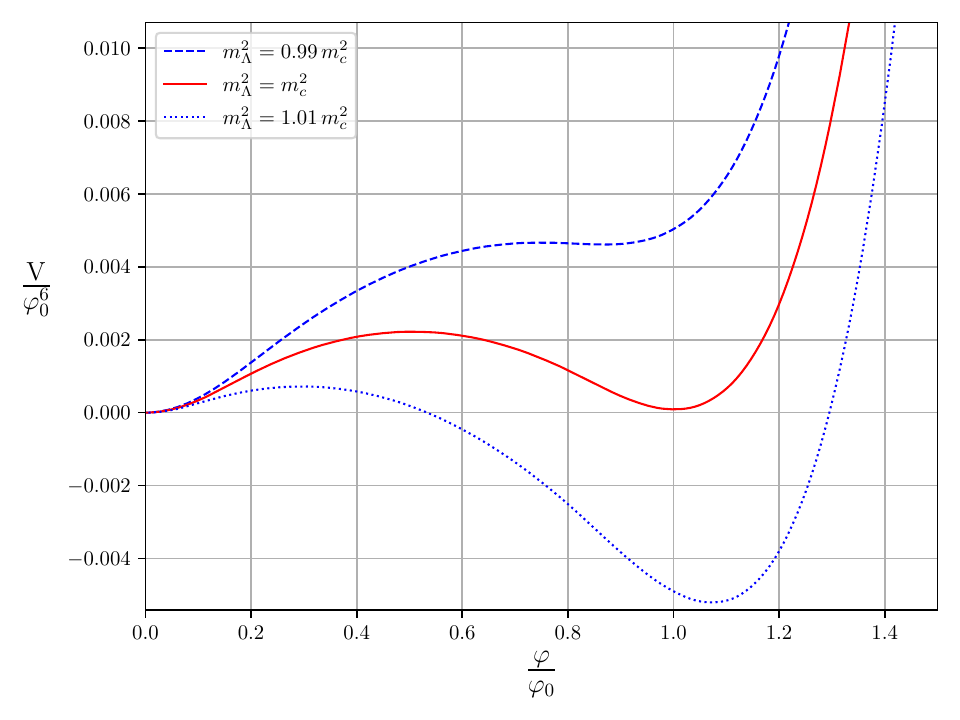}
\caption{Thermodynamics potential above (dashed line) and below (dotted line) the first-order phase transition for $n=4$.}
\label{fig:ir}
\end{figure}

To find how the potential acquires a new non trivial minimum and under what conditions the solution $\varphi_0$ does correspond to a stable phase, the RG flow should be terminated at some infrared scale $k_{IR}$ determined by the specific problem under consideration,  a deeper scale $k < k_{IR}$ leads to nothing but false precision, see Appendix~\ref{sec:numer}.

\begin{figure}[t!]
\includegraphics[width=0.48\textwidth]{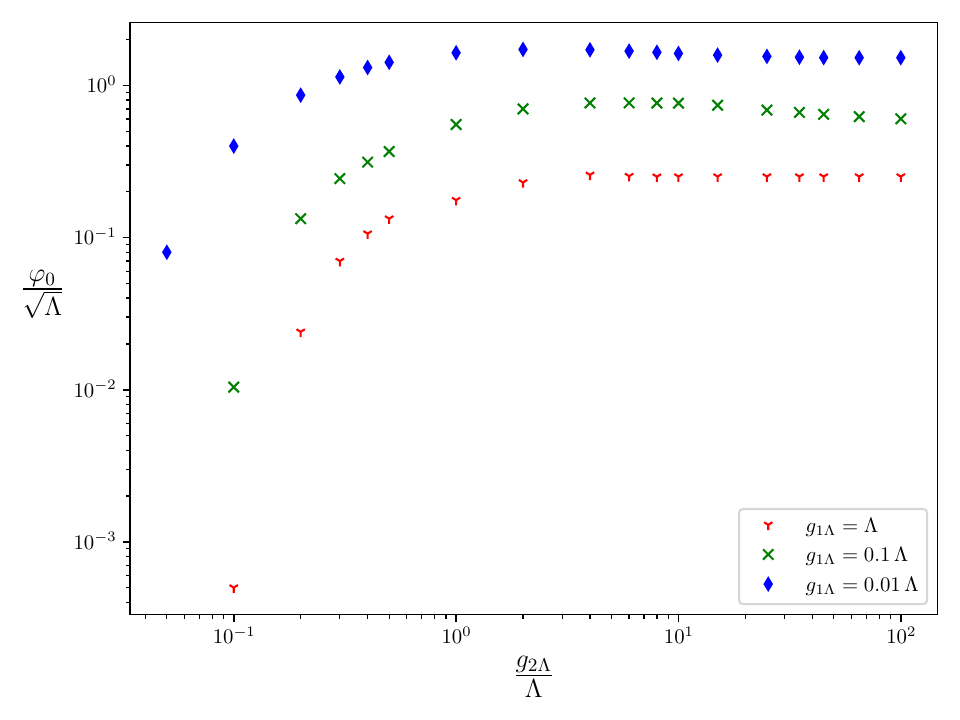}
\caption{Finite jump in the order parameter at the phase transition point $m^2_c$ for $n=4$.  For large magnitudes of $g_{2 \Lambda}$, the system settles down to the strong first-order transition which is specified by the  jump plateaus.}
\label{fig:phi0}
\end{figure}

\begin{figure}[t!]
\includegraphics[scale=0.5]{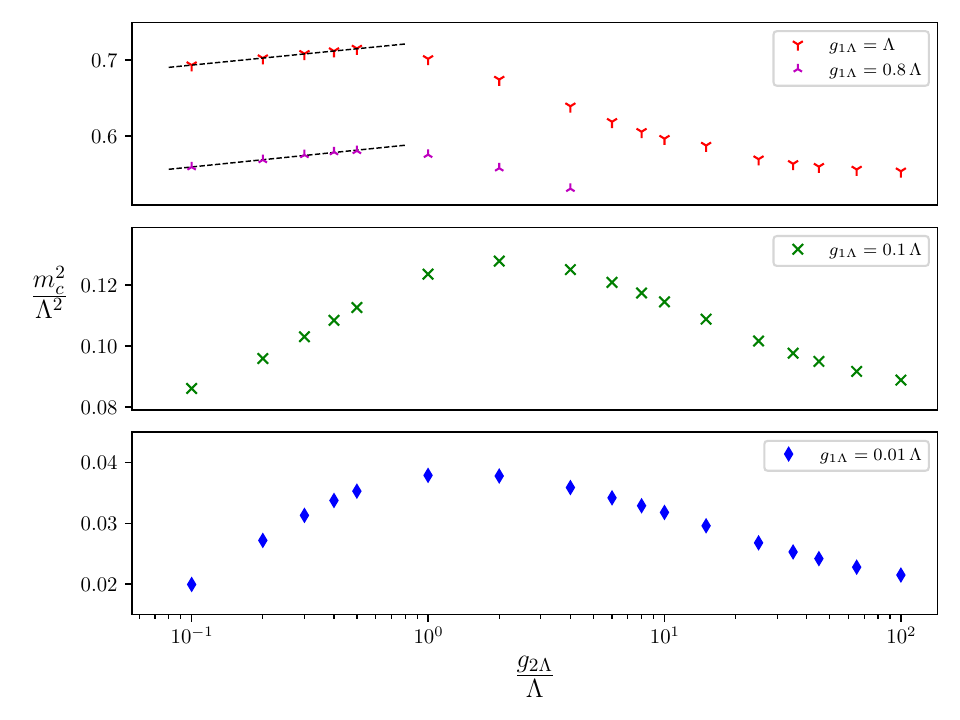}
\caption{The critical value of the ultraviolet parameter $m^2_\Lambda$ at which the first-order transition takes place for $n=4$. In the weak first-order transition regime the curves are linear  with the same slope.}
\label{fig:mc}
\end{figure}

In the case of present interest, that of a discontinuous phase transition, we will immediately
restrict our search to the stable state deﬁned by the conventional conditions: $\partial \vv /\partial \varphi|_{\varphi = \varphi_0} = 0$, $ \vv(0) = \vv(\varphi_0) $ and finally $\partial^2 \vv /\partial \varphi^ 2|_{\varphi = \varphi_0} > 0$, where $V$ denotes the infrared potential $V_{k = k_{IR}}$. The ultimate shape of the potential $V$ certainly depends on the values of the ultraviolet parameters $m^2_{\Lambda}$, $g_{1 \Lambda}$, and $g_{2 \Lambda}$. Among them, for continuous phase transitions, the relevant parameter $m^2_{\Lambda}$ is considered to be the measure of the distance to the phase transition or critical point, while the specific values of $g_{1 \Lambda}$ and $g_{2 \Lambda}$ are usually unimportant. However, the charge $g_{2 \Lambda}$  acts now as a factor that drives the first-order phase transition, the evolution of which  is shown in  \cref{fig:evolution}. Along with the emergence of the double-well structure of the potential $\vv$, the next invariant expansion coefficient $\ww$, being a constant in the ultraviolet scale, starts to develop a nontrivial shape. For  given $g_{1 \Lambda}$ and $g_{2 \Lambda}$ parameters, one can find the system in ordered or disordered phases depending on the specific value of $m^2_{\Lambda}$, as shown in \cref{fig:ir}. The value of the order parameter gap at the phase transition point $m^2_c$ is defined more by the magnitude of $g_{2 \Lambda}$, see \cref{fig:phi0}, and one can distinctly identify weak $g_{2 \Lambda}/\Lambda < 1$ and strong $g_{2 \Lambda}/\Lambda \gg 1$ regimes. In the former case, the jump is small enough and the first-order transition picture becomes less visible. As we decrease $g_{2 \Lambda}$ to zero, the value of the order parameter decreases and eventually tends to zero at a second-order phase transition. In contrast, the strong regime is characterized by a plateau, where the finite jump $\varphi_0$ ceases to depend on the specific amplitude of $g_{2 \Lambda}$. We also analyzed the phase transition value  $m^2_c$ as a function of the coupling constants, \cref{fig:mc}. This quantity as a rule is not universal, but in the weak regime where $g_{2 \Lambda} < g_{1 \Lambda}$ the respective curve  reveals a small constant slope, which does not depend on the given value of  $g_{1 \Lambda}$ and, in this sense, the behavior of the physical system can be considered to be pseudo-universal.

\section{\label{sec:end}Conclusions}
In this work, we have discussed possible phase transitions in the matrix field model in the context of $3D$ superfluidity in the multicomponent $SU(n)$ Fermi system. To this end, we have computed a full thermodynamic potential within the local potential approximation of the functional renormalization group approach. This allows us to find that, although a tree-approximation for the potential predicts a continuous phase transition to superfluid phase, fluctuations captured by RG induce a discontinuous phase transition in a system with $n>2$ spin degrees of freedom per a particle. In the vicinity of the upper critical dimension $d = 4 - \varepsilon$, our findings are directly in line with known perturbative renormalization group results achieved within the $\varepsilon$-expansion. Besides the routine relevant parameter, $m^2_\Lambda$ in our notation, the coupling $g_{2\Lambda}$ turns out to be a control variable. It determines the strength of the obtained discontinuous phase transition, which encompasses weak and strong  regimes corresponding to small and large magnitudes of $g_{2\Lambda}$, respectively. In case the ultraviolet bosonic theory is deduced from the original fermionic action, these quantities are evaluated as functions of temperature, chemical potential, and scattering length, and hence can be tuned to bring the system into a particular regime. 
Close to the weak transition the jump of the superfluid density is barely discernible. Consequently, the system will apparently manifest continuous behavior in experimental reality, while the  non-universal nature of the first-order transition becomes more pronounced in the strong regime. The generic structure of the established pattern persists for all $n>2$ up to the limit $n \to \infty$, where the flow equations turn into those explored previously in \cite{kalagov2020}.  

We obtained the present findings for the fully bosonised effective action \cref{eq:GL}. Regarding the follow-up application to physics of ultracold multicomponent systems, e.g. BCS-BEC crossover, few-body phenomena, unitary gas, it would also be reasonable to consider a partially bosonised model, where the quartic fermionic interaction was decoupled via the Hubbard-Stratonovich trick, with subsequent simultaneous integration of  fermionic and bosonic modes out within the FRG on the basis of an appropriate truncation for the effective average action \cite{Faigle2021,Boettcher2012,Diehl2010,Roscher2015}. 
There is also room to improve on the numerical computational techniques using the aforementioned discontinuous methods. The performed analysis can be surely extended to these issues, and it therefore provides an avenue for forthcoming theoretical revealing the relationship between large spin degrees of freedom and superfluidity/superconductivity in a fermionic matter. 

\begin{acknowledgments}
We thank Jan M. Pawlowski for thoughtful discussions of the FRG machinery. Georgii Kalagov was supported by the Foundation for the Advancement of Theoretical Physics
and Mathematics ``BASIS''. Michal Hnati\v{c} was supported by the Ministry of Education, Science, Research and Sport of the Slovak Republic(VEGA Grant No. 1/0535/21).

\end{acknowledgments}

\appendix
\section{\label{sec:mass}Mass spectrum}
To obtain the mass spectrum, we need to shift the field by constant background $\Phi \to \Phi_0 +\Phi$ and then diagonalize the respective quadratic form. To perform this step, we represent the complex fluctuation field $\Phi$ in the form $\Phi = \phi^{(1)} + i \,\phi^{(2)}$ with the real skew-symmetric matrices $\phi^{(1)}, \phi^{(2)}$:
\begin{equation}
\phi^{(j)} = \begin{pmatrix}
 A^{(j)} + C^{(j)} & B^{(j)}_a + B^{(j)}_s \\ 
 B^{(j)}_a-B^{(j)}_s & A^{(j)} - C^{(j)} \\
 \end{pmatrix} +  \zeta^{(j)}_1 \begin{pmatrix}
 0 & I_{n/2} \\ 
-I_{n/2} & 0 \\
 \end{pmatrix}, 
\end{equation}
 the superscript $j = 1, 2$, and  $A^{(j)}, B_a^{(j)}, C^{(j)}$ -- $(n/2)\times(n/2)$-antisymmetric matrices; while $B^{(j)}_s$ -- $(n/2)\times(n/2)$-symmetric ones. We represent  each of these matrices in the following form: 
\begin{align*}
A^{(j)} &= \begin{pmatrix}
 \zeta^{(j)}_2\, J & A^{(j)}_1 \\ 
 -{A^{(j)}_1}^{T} & A^{(j)}_2 \\
 \end{pmatrix}, \,
C^{(j)} = \begin{pmatrix}
 \zeta^{(j)}_3 \, J & C^{(j)}_1 \\ 
 -{C^{(j)}_1}^{T} & C^{(j)}_2 \\
 \end{pmatrix}, \, \\[1ex]
B^{(j)}_s &= \begin{pmatrix}
 B^{(j)}_0 + \frac{1}{2}\, \zeta^{(j)}_4 \, I_2 & B^{(j)}_1 \\ 
 {B^{(j)}_1}^{T} & B^{(j)}_2 - \frac{2}{n-4}\, \zeta^{(j)}_4 \, I_{n/2-2} \\
 \end{pmatrix},\, \\[1ex]
B^{(j)}_a &= \begin{pmatrix}
  \zeta^{(j)}_5 \, J & B^{(j)}_3 \\ 
 -{B^{(j)}_3}^{T} & B^{(j)}_4\\
 \end{pmatrix},
\end{align*}
where $A^{(j)}_1, C^{(j)}_1, B^{(j)}_1, B^{(j)}_3$ -- $(2) \times (n/2-2)$-matrices, for these matrices we introduce the two component row $A^{(j)}_1 = ( A^{(j)}_{11}, A^{(j)}_{12} )^T$ made of $(n/2-2)$-dimensional vectors $A^{(j)}_{11}$ etc., they will be called below vector modes; $\zeta^{(j)}_1,\dots,\zeta^{(j)}_5$ -- scalar modes; $A^{(j)}_2, C^{(j)}_2, B^{(j)}_4$ -- $(n/2-2) \times (n/2-2)$-antisymmetric matrix modes; $B^{(j)}_0$ and $B^{(j)}_4$ -- traceless symmetric matrix modes sizes of which are $(2) \times (2)$ and $(n/2-2) \times (n/2-2)$, respectively. Diagonalization procedures yield the eigenvalue degeneracies $n_{\text{mode}}$, which satisfy the condition $\sum_{\text{all modes}} n_{\text{mode}} = n (n - 1)$, and the eigenvalues listed bellow.
\begin{widetext}
\begingroup
\allowdisplaybreaks
\begin{align*}
&\text{Real scalar masses ($j = 1$):}\\[1ex]
M^2_{\zeta,1} & = M^2_{\zeta,2} =  \V_k' +  \left(\W_k'+ \frac{n-2}{2 \rho_1} \W_k \right) \rho_2,\\
M^2_{\zeta,3} & =  \V_k' + 2 \,{ \V_k''}\,  \rho_{1} +  \left(2\,{ \W''_k}\,\rho_{{1}}+5\,{ \W_k'}+{\frac {\W_k}{\rho_{{1}}}} -\frac {n \V_k''}{2 \rho_{1}}
 \right)\rho_{{2}} +\,{\frac {\rho_{{2}}
 \left( n{\V_k''}+4\,{\W_k'}\,\rho_{{1}}+4\,\W_k \right) ^{2}}{ 2 \rho_{1} \left( n
{\V_k''}-2\,\W_k \right)}},\\
M^2_{\zeta,4} &= \V'_{k} + \frac{4 \V_k}{n} \rho_1 + 3\sqrt{\frac{ n - 4}{n}} \W_k \sqrt{\rho_2}+ \left( 3 \W'_k +  \frac{3 n-14}{4 
\rho_1}\W_k  + \frac{n \V''_k}{4 \rho_1} \right)\rho_2 -\,{\frac {\rho_2
 \left( n \V''_k+4\,\W'_k\,\rho_1+4\,\W_k\right) ^{2}}{ 4 \rho_1\left( n
\V''_k-2\,\W_k \right)}},\\
M^2_{\zeta, 5} &= \V'_{k} + \frac{4 \V_k}{n} \rho_1 - 3\sqrt{\frac{ n - 4}{n}} \W_k \sqrt{\rho_2}+ \left( 3 \W'_k +  \frac{3 n-14}{4 
\rho_1}\W_k  + \frac{n \V''_k}{4 \rho_1} \right)\rho_2 -\,{\frac {\rho_2
 \left( n \V''_k+4\,\W'_k\,\rho_1+4\,\W_k\right) ^{2}}{ 4 \rho_1\left( n
\V''_k-2\,\W_k \right)}}. \\[1ex]
%\end{align*}
&\text{Imaginary scalar masses ($j = 2$):} \\[1ex]
%\begin{align*}
M^2_{\zeta,1} &= \V_k' +  \left(\W_k'- \frac{\W_k}{\rho_1} \right) \rho_2, \\
 M^2_{\zeta,2} &= M^2_{\zeta,3} = \V_k' + \frac{4  \W_k}{n} \rho_1 +  \left( \W_k'+ \frac{n -6}{2 \rho_1} \W_k \right) \rho_2 , \\
 M^2_{\zeta,4} &=\V_k' + \W_k \sqrt{\rho_2} + \left( \W_k'+ \frac{n -4}{4 \rho_1} \W_k \right) \rho_2,\\
  M^2_{\zeta,5} &=\V_k' - \W_k \sqrt{\rho_2} + \left( \W_k'+ \frac{n -4}{4 \rho_1} \W_k \right) \rho_2. \\[1ex]
&\text{Real vector masses ($j = 1$): $n_v = n/2-2$} \\[1ex]
M^2_{v, 1} &= M^2_{v, 2} =  \V_k' + \frac{\W_k}{2} \sqrt{\rho_2} + \left( \W_k'+ \frac{3 n -16}{16 \rho_1} \W_k \right) \rho_2, \\
M^2_{v, 3} &= M^2_{v, 4} = \V_k' - \frac{\W_k}{2} \sqrt{\rho_2} + \left( \W_k'+ \frac{3 n -16}{16 \rho_1} \W_k \right) \rho_2, \\    
M^2_{v, 5} & = M^2_{v, 6} = \V_k' +  \frac{4 \W_k}{n} \rho_1+ \frac{3 \W_k}{2}  \sqrt{\rho_2} + \left( \W_k'+ \frac{5 n -48}{16 \rho_1} \W_k \right) \rho_2,\\
M^2_{v, 7} &=M^2_{v, 8}  = \V_k' + \frac{4 \W_k}{n} \rho_1 - \frac{3 \W_k}{2}  \sqrt{\rho_2}+ \left( \W_k'+ \frac{5 n -48}{16 \rho_1} \W_k \right) \rho_2. \\[1ex]
&\text{Imaginary vector masses ($j = 2$): $n_v = n/2-2$} \\[1ex]
M^2_{v, 1} &= M^2_{v, 2} =  \V_k' + \frac{\W_k}{2} \sqrt{\rho_2} + \left( \W_k'+ \frac{3 n -16}{16 \rho_1} \W_k \right) \rho_2, \\
M^2_{v, 3} &= M^2_{v, 4} = \V_k' - \frac{\W_k}{2} \sqrt{\rho_2} + \left( \W_k'+ \frac{3 n -16}{16 \rho_1} \W_k \right) \rho_2, \\    
M^2_{v, 5} & = M^2_{v, 6} = \V_k' +  \frac{4 \W_k}{n} \rho_1+ \frac{3 \W_k}{2}  \sqrt{\rho_2} + \left( \W_k'+ \frac{5 n -48}{16 \rho_1} \W_k \right) \rho_2,\\
M^2_{v, 7} &=M^2_{v, 8}  = \V_k' + \frac{4 \W_k}{n} \rho_1 - \frac{3 \W_k}{2}  \sqrt{\rho_2}+ \left( \W_k'+ \frac{5 n -48}{16 \rho_1} \W_k \right) \rho_2. \\[1ex]
&\text{Real matrix masses ($j = 1$):} \\[1ex]
  M^2_{B_0} & =  \V_k' + \frac{4  \W_k}{n} \rho_1 +  \left( \W_k'+ \frac{n -6}{2 \rho_1} \W_k \right) \rho_2, \quad n_{B_0} = 2,\\
   M^2_{B_2} & =  \V_k' + \frac{4  \W_k}{n} \rho_1 +  \left( \W_k' - \frac{3}{\rho_1} \W_k \right) \rho_2, \quad n_{B_2} = (n/2-2) (n/2-1)/2 - 1,\\
    M^2_{A_2} & =  \V_k' + \frac{4  \W_k}{n} \rho_1 +  \left( \W_k' - \frac{3}{\rho_1} \W_k \right) \rho_2, \quad n_{A_2} = (n/2-3) (n/2-2)/2,\\
    M^2_{B_4} & = M^2_{C_2} = \V_k'  +  \left( \W_k' - \frac{1}{\rho_1} \W_k \right) \rho_2, \quad n_{B_4} = n_{C_2} = (n/2-3) (n/2-2)/2. \\[1ex]
&\text{Imaginary matrix masses ($j = 2$):} \\[1ex]
  M^2_{B_0} & =  \V_k' +   \left( \W_k'+ \frac{n - 2}{2 \rho_1} \W_k \right) \rho_2, \quad n_{B_0} = 3,\\
   M^2_{B_2} & =  \V_k'  +  \left( \W_k' - \frac{1}{\rho_1} \W_k \right) \rho_2, \quad n_{B_2} = (n/2-2) (n/2-1)/2 - 1,\\
    M^2_{A_2} & =  \V_k' +  \left( \W_k' - \frac{1}{\rho_1} \W_k \right) \rho_2, \quad n_{A_2} = (n/2-3) (n/2-2)/2,\\
    M^2_{B_4} & = M^2_{C_2}=  \V_k'  +\frac{4  \W_k}{n} \rho_1 + \left( \W_k' - \frac{3}{\rho_1} \W_k \right) \rho_2, \quad n_{B_4} = n_{C_2} = (n/2-3) (n/2-2)/2.
\end{align*}
\endgroup
\end{widetext}
Finally, one can derive the desired  RG flow by equating the coefficients of the corresponding powers of $\rho_2$
 \begin{align*}
      \partial_k  \V_k(\rho_1) + \rho_2  \partial_k  \W_k(\rho_1) &=  C_d  k^{d+1} \sum_a \frac{n_a}{k^2 + M_a^2(\rho_1, \rho_2)}\\
      &\approx \flow\{ \V_k\} + \rho_2 \, \flow\{ \W_k\},
 \end{align*}
where $C_d^{-1} = (4 \pi)^{d/2} \Gamma(d/2+1)$. This immediately leads to the system presented in the main text. 

To derive the large $n$ expansion, one needs to additionally rescale the computed observables and eliminate the leading $n$-dependency. We will then properly pass to the limit $n \to \infty$.  The skew-symmetric complex field can be expanded into the $SO(n)$ generators with $n(n-1)$ real coefficients; so we expect $\rho_1 \sim n^2$ as $n\to \infty$ and introduce, therefore, a new variable $\rho_1 \to n^2 \rho_1$, while the potential functions and their derivatives are scaled as
\begin{align*}
   &\V_k \to n^2 \V_k, \quad \V_k' \to \V_k',\quad  \V_k'' \to  \V_k''/n^2,\\[1ex]
   &\W_k \to \W_k/n, \quad  \W_k' \to \W_k'/n^3, \quad  \W_k'' \to  \W_k''/n^5.
 \end{align*}
The leading in $1/n$ contribution to the flow is then cast in the form
\begin{align}
\nonumber
  \partial_k  &\vv_k(\rho_1) =   C_d\,  k^{d+1} \left\{ \frac{1/2}{k^2+\vv_k' + 4 \rho_1 \ww_k} + \frac{1/2}{k^2+\V_k'}  \right\},\\[1ex]  \nonumber
  \partial_k & \ww_k(\rho_1) =   C_d\,  k^{d+1} \left\{ \frac{ \ww_k^2}{(k^2+\vv_k')^3}  - \frac{(\ww_k + 2 \, \rho_1 \ww_k')}{
 4 \rho_1 (k^2+\vv_k')^2} \right. \\[1ex]  \nonumber 
 &+ \frac{9 \ww_k^2}{(k^2+\vv_k' + 4 \rho_1 \ww_k)^3}    +  \left.\frac{(\ww_k - 2 \, \rho_1 \ww_k')}{ 4 \rho_1 (k^2+\vv_k' + 4 \rho_1 \ww_k)^2}  \right\}.
\end{align}
The obtained system coincides with its counterpart  derived previously in \cite{kalagov2020} for the adjoint representation of $SU(n)$.   In that work, it was shown that in the infrared limit fluctuations manifest themselves in the emergence
of the typical for the first-order transition shape of the potential $\vv_k(\rho_1)$. Qualitative behavior of the phase transition parameters such as $m_c^2, \phi_0$ akin to the picture elucidated above for $n>2$.

\section{\label{sec:numer}Numerical integration}
The first step towards obtaining a numerical solution of the PDEs is to construct a spatial discretization on the appropriate grid. The obtained system of ordinary differential equations is then amenable for future treatment within the adaptive Runge-Kutta scheme. 

In this paper, we will focus on the  collocation method \cite{Boyd1989}, where the Chebyshev-Lobatto collocation points $x_{k, N} = \cos(\pi k/N), \, k = 0\dots N$ are the extrema of the $N$-th order Chebyshev polynomial $T_N(x) = \cos(N \arccos{(x)})$; thus the approximation of the smooth solution $f(x, t)$ in the domain $x \in [-1, 1]$ reads 
\begin{equation} \notag
    f(x, t) \approx \sum\limits_{k = 0}^{N}  f_{k,N} \, h_k(x), \quad f_{k,N} = f(x_{k, N}, t).
\end{equation}
The cardinal functions $h_k(x)$ being the Lagrange interpolation polynomials of degree $N$ meet the condition $h_k(x_i) = \delta_{k i}$ and are defined by 
\begin{equation} \notag
 h_k(x) = \frac{(-1)^{k+1}(1-x^2) T'_N(x)}{c_k N^2 (x - x_{k, N})}, 
\end{equation}
where $c_0 = c_N = 2$ and $c_k = 1$ for $k = 1\dots N-1$. The spatial derivatives of the $M$-th order of the function at the collocation points are immediately evaluated and reduced to the matrix-vector multiplication $f^{(M)}_{k,N}= (D^{M})_{k j} f_{j,N}$; here the derivative matrix $D_{k j} = h'_j(x_{k, N})$ has the explicit form
\begin{align}\notag
    &D_{k j} = \frac{c_k}{c_j}  \frac{(-1)^{k+j}}{(x_{k, N} - x_{j, N})} \;\; \text{for} \;\;  k \neq j; \\  \notag
    & D_{k k} = -\frac{1}{2}  \frac{x_{k, N}}{(1 - x_{k, N}^2)} \;\; \text{for} \;\; k \neq 0, N; \\ \notag
    & D_{0 0} = - D_{N N} = \frac{2 N^2 + 1}{6}.
\end{align}
The initial value problem for the partial differential equation 
\begin{equation} \notag
\partial_t f = \mathcal{F}(f, f', f''), \quad f(x, 0) = \psi(x), \, f(1, t) = \psi(1),
\end{equation}
is then  reduced to the system of $N$ ordinary differential equations 
\begin{align} \notag
    \partial_t  f_{k, N} = \mathcal{F}(f_{k,N},\, f^{(1)}_{k,N}, \,  f^{(2)}_{k,N}),
    \quad  f(x_{k, N}, 0) = \psi(x_{k, N}),
\end{align}
with the boundary condition $f(x_{0, N}, t) = \psi(1)$ and $ k = 1,\dots,N$.

 \begin{figure}[t!]
\includegraphics[scale=0.5]{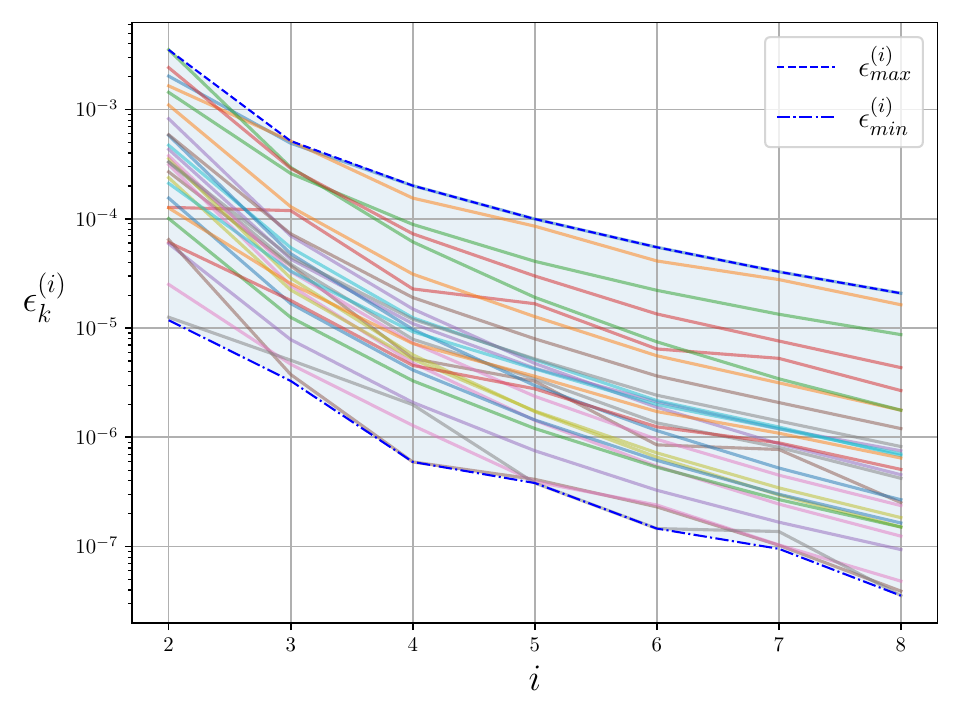}
\caption{The convergence of the numerical scheme with respect to the number of nodes $N_i = 30 \,(i-1)$ for the potential presented in \cref{fig:ir}, $k_{IR} \approx 0.05 \Lambda$. 
The family of curves squeezed between boundaries $\epsilon^{(i)}_{max}$ and $\epsilon^{(i)}_{min}$ depicts
the relative errors at the $k$-th nodes.}
\label{fig:error}
\end{figure}

At the given nodes number $N$ the solution of the system, corresponding to the flow,  \cref{eq:rgeq}, approaches the fixed value as the scale $k$ decreases and the flow significantly slows down; therefore, one can stop all subsequent Runge-Kutta steps at some value $k_{IR}$. We solve the system with the relative accuracy of the order $10^{-3}$ for any initial data, thus the quantity $k_{IR}$ is chosen so that the relative change in solution for scales $k < k_{IR}$ is less than $10^{-3}$. Depending on the initial values of the potentials, $k_{IR}$ should be taken appropriately, and in our numerical computation the characteristic values of $k_{IR}$ lie in the range  $\sim 0.1 - 0.05$. 

To verify numerical convergence rates  at the infrared point $k_{IR}$, we sequentially increase the number of collocation points $N = 30, 60,120,\dots,240$, and evaluate the relative differences  $\epsilon^{(i)}_k = |f_{k, 30 (i-1)}-f_{k, 30 i}|/|f_{k, 30 i}|$ at the $k$-th node, where $i = 2,\dots,8$. The upper estimation  $\epsilon^{(i)}_{max}$ at the $i$-th step is given by the maximum value of the $\epsilon^{(i)}_{k}$ across all nodes. Figure \ref{fig:error} shows the convergence of the employed numerical scheme with respect to the number of nodes. The used numbers of collocation points $N$ is typically $\sim 150 - 200$, being enough to ensure the desired total accuracy $10^{-3}$.

\bibliography{sun}
\end{document}